\documentclass{jpsj-suppl}
\usepackage{color}

\inst{$^{1}$Institute of Modern Physics, Chinese Academy of
Sciences, Lanzhou 730000, China \\
$^{2}$ University of Chinese Academy of Sciences, Beijing 100049, China \\
$^{3}$ Joint Institute for Nuclear Research, Dubna 141980, Russia}
\email{xywang@impcas.ac.cn;\\
 avg@jinr.ru}
\recdate{Jan  XX, 2016}
\abst{ Photoproduction of the charmonium-like state $Z_{c}(4200)$ and the charmed baryon $\Lambda _{c}^{\ast }(2940)$ is investigated with an effective Lagrangian approach and the Regge trajectories applying to the COMPASS experiment. Combining the experimental data from  COMPASS and our theoretical model we estimate the upper limit
of $\Gamma _{Z_{c}(4200)\rightarrow J/\psi \pi }$ to be of about 37 MeV. Moreover, the possibility to  produce $\Lambda _{c}^{\ast }(2940)$ at COMPASS is discussed. It seems one can try to search for this hadron in the
missing mass spectrum since the $t$-channel is dominating for the $\Lambda _{c}^{\ast }(2940)$ photoproduction.}
\kword{charmed hadron, photoproduction, COMPASS}

\begin{document}

\title{Charmed hadron photoproduction at COMPASS}
\author{Xiao-Yun Wang$^{1,2}$ and Alexey Guskov$^{3}$}
\maketitle

\section{Introduction}

In the past decades {the} great progress has been achieved in {the} hadron
spectroscopy, {largely} due to the high quality photoproduction data
obtained at electronmagnetic facilities. {Recently a series of exotic states
{was} found in the experiments greatly promotes} studying of the inner
structure of hadrons and the strong interaction \cite{lx14,sl15}.
Photoproduction provides {one of the most} direct routes to information
about {the} hadronic structure. The COMPASS experiment at CERN, {running
since 2002 and using} a positive muon beam of 160 GeV/c (2002--2010) or 200
GeV/c momentum (2011), {is the} ideal platform for studying the
photoproduction of particle at high energies since it covers the range of $W$
up to 19.4 GeV \cite{compass,a2xy}. {The photoproduction processes at high
energies can be described effectively} with Regge-trajectory exchanges.

In 2014 a new charged charmonium-like $Z_{c}^{+}(4200)$ was observed in the
invariant mass spectrum of $J/\psi \pi ^{+}$ with a significance of 6.2$%
\sigma $ by {the} Belle Collaboration \cite{belle14}. The relevant
experiment results show that the spin-parity quantum numbes of $%
Z_{c}^{+}(4200)$ {corresponds to} $1^{+}$. Furthermore, its mass and width
are measured to be \cite{belle14}
\begin{eqnarray*}
M &=&4196_{-29-13}^{+31+17}\text{ MeV}, \\
\Gamma  &=&370_{-70-132}^{+70+70}\text{ MeV.}
\end{eqnarray*}%
The charmonium-like $Z_{c}^{+}(4200)$ {state is very broad} and its inner
structure arouse great {interest} among the researchers. Within the
different theoretical {frames the $Z_{c}^{+}(4200)$} was explained as the
tetraquark or {the} molecule states etc. \cite{zsl14,wei11,wei15,wzg15}.
These experimental and theoretical results indicate that the $Z_{c}^{+}(4200)
$ is the ideal candidate for {investigation} and understanding the nature of
exotic charmonium-like states.

The charmed baryon $\Lambda _{c}^{\ast }(2940)$ was first {discovered} by
the \textit{BABAR} Collaboration \cite{babar07} by analyzing the $pD^{0}$
invariant mass spectrum. Later, the Belle Collaboration \cite{belle07}
confirmed it as a resonant structure in the final state $\Sigma
_{c}(2455)^{0,++}\pi ^{\pm }\rightarrow \Lambda _{c}^{+}\pi ^{+}\pi ^{-}$.
The values for the mass and width of the $\Lambda _{c}^{\ast }(2940)$ state
were reported by both Collaborations \cite{babar07,belle07}, which are
consistent with each other:%
\begin{eqnarray*}
\text{\textit{BABAR }}\text{: } &M&=2939.8\pm 1.3\pm 1.0\text{ MeV,} \\
&&\Gamma =17.5\pm 5.2\pm 5.9\text{ MeV,} \\
\text{Belle}\text{: } &M&=2938.0\pm 1.3_{-4.0}^{+2.0}\text{ MeV,} \\
\text{ \ \ \ \ } &&\Gamma =13_{-5-7}^{+8+27}\text{ MeV.}
\end{eqnarray*}

However, the spin-parity of the $\Lambda _{c}^{\ast }(2940)$ state have not
been determined {yet in experiments}. Since the nature of $\Lambda
_{c}^{\ast }(2940)${\ still is unclear, an additional efforts are} needed to
determine its real inner structure. As of now, all experimental observations
of the $\Lambda _{c}^{\ast }(2940)$ have been{\ performed in} the $e^{+}e^{-}
$ collisions \cite{babar07,belle07}. Since there is no any experimental
information {for} the photoproduction of the $\Lambda _{c}^{\ast }(2940)$,
it is highly necessary to give a study on the photoproduction of{\ this
hadronic sate}.

In this paper, we give a compact review of {our} theoretical results for the
photoproduction of the $Z_{c}(4200)$ and $\Lambda _{c}^{\ast }(2940)$. In
the next section we discuss the {photoproduction of $Z_{c}(4200)$} and the
partial decay width $\Gamma _{Z(4200)\rightarrow J/\psi \pi }.$ In {the}
Section III the results {for} $\Lambda _{c}^{\ast }(2940)$ production in {%
the reaction $\gamma n\rightarrow D^{-}\Lambda _{c}^{\ast }(2940)^{+}$ are
presented.} Finally this paper concludes with a short summary.

\section{$Z_{c}(4200)$ photoproduction}

In order to describe the production of $Z_{c}(4200)$ in {the} $\gamma
p\rightarrow Z_{c}^{+}(4200)n$ reaction via {the} pion exchange, we
introduce {the} pion Reggeized treatment by replacing the Feynman propagator
{by} the Regge propagator (more details can be found in the Ref. \cite{z4200}%
). Fig. 1 presents the total cross section $\sigma (\gamma p\rightarrow
Z_{c}^{+}n)$ through {the} $\pi $ meson or pionic Regge trajectory exchange
by taking $\Lambda =0.7$ GeV. It is found that Reggeized treatment can lead
to a lower cross section of the $Z_{c}^{+}(4200)$ photoproduction at high
photon energies. Moreover, one {can note} that the peak position of {the}
total cross section was moved to the higher {energies} when the Reggeized
treatment is considered.

{The $J/\psi \pi ^{\pm }$ mass spectrum presented by the COMPASS
collaboration in \cite{compass} does not exhibit any statistically
significant structure at about $4.2$ GeV (see Fig. 2). However, we can use
this experimental result to estimate an upper limit for the value $%
BR(Z_{c}(4200)\rightarrow J/\psi \pi )\times \sigma _{\gamma N\rightarrow
Z_{c}(4200)N}$.}

With the normalization used in \cite{compass} this limit corresponds to the
result
\begin{equation}
BR(Z_{c}(4200)\rightarrow J/\psi \pi )\times \sigma _{\gamma N\rightarrow
Z_{c}(4200)N}<340~\mathrm{pb}.
\end{equation}

It can be used for estimation of an upper limit for the partial width $%
\Gamma _{J/\psi \pi }$ of the decay $Z_{c}({4200)}\rightarrow J/\psi \pi $
combined with our theoretical result. The production cross section, averaged
over the $W$-range covered by COMPASS, is about $\Gamma _{J/\psi \pi }\times
91$~pb/MeV. Then
\begin{equation}
\frac{\Gamma _{J/\psi \pi }}{\Gamma _{tot}}\times \sigma _{\gamma
N\rightarrow Z_{c}^{\pm }(4200)~N}=\frac{\Gamma _{J/\psi \pi }^{2}\times 90~%
\mathrm{pb/MeV}}{\Gamma _{tot}}<340~\mathrm{pb}.
\end{equation}%
Assuming $\Gamma _{\mbox{tot}}=370$~MeV, one obtain an upper limit $\Gamma
_{J/\psi \pi }<37$ MeV.
\begin{figure}[h]
\centering
\includegraphics[scale=0.4]{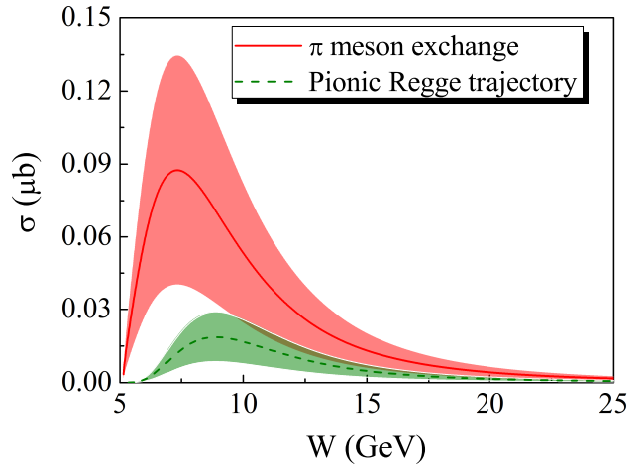}
\caption{(Color online) The total cross section of the $\protect\gamma %
p\rightarrow Z_{c}^{+}(4200)n$ reaction through $\protect\pi $ meson or
pionic Regge trajectory exchange. Here, the numerical results are calculated
with $\Gamma _{Z_{c}(4200)\rightarrow J/\protect\psi \protect\pi }=87.3$
MeV, while the bands stand for the uncertainties with the variation of $%
\Gamma _{Z_{c}(4200)\rightarrow J/\protect\psi \protect\pi }$ from 40.2 to
131.4 MeV.}
\end{figure}

\begin{figure}[tb]
\centering
\includegraphics[scale=0.4]{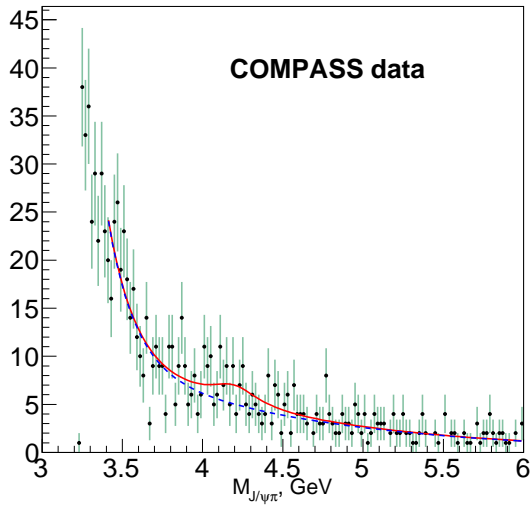}
\caption{(Color online) Mass spectrum of the $J/\protect\psi \protect\pi%
^{\pm} $ state obtained by COMPASS \protect\cite{compass}. The fitted
function is shown as a red solid line. Dashed blue line corresponds to the
continuum description.}
\end{figure}

\section{$\Lambda _{c}^{\ast }(2940)$ photoproduction}

We also studied the production of {the $\Lambda _{c}^{\ast }(2940)$ state}
in the $\gamma n\rightarrow D^{-}\Lambda _{c}^{\ast +}$ process (more
details can be found in Ref. \cite{lam2940}). Fig. 3 shows that the total
cross section for {the} $\gamma n\rightarrow D^{-}\Lambda _{c}^{\ast +}$
process for {two sets of $J^{P}$: $\frac{1}{2}^{+}$ and $\frac{1}{2}^{-}$}.
One {can notice} that the contribution from the $t$-channel with $D^{\ast }$
exchange is dominant, while the contributions from $t$-channel {with the} $D$
meson, $s$-channel {with the} nucleon pole, $u$-channel {with the} $\Lambda
_{c}^{\ast }$ exchange and {the} contact term are small.

Basing on the integrated luminosity and the calculated $\Lambda _{c}^{\ast
}(2940)$ production cross section value of 0.02 $\mu $b ($J^{P}=\frac{1}{2}%
^{+}$, $\Lambda $=3.0 GeV, $\Gamma _{\Lambda _{c}^{\ast }\rightarrow
pD^{0}}=0.21$ MeV) one can expect to find in the COMPASS muon data sample
collected between 2002 and 2011 up to 0.9$\times 10^{5}$ $\Lambda _{c}^{\ast
}(2940)$ baryons produced via the reaction $\gamma n\rightarrow D^{-}\Lambda
_{c}^{\ast +}$. This number can be compared with the COMPASS open charm
leptoproduction results based on the data collected between 2002 and 2007
\cite{compassoch} where the number of reconstructed $D^{0}\rightarrow
K^{+}\pi ^{-}$ decays (BR=3.88\%) exceeded $5\times 10^{4}$.

Since the t-channel is dominating, the energy transferred to the produced $%
\Lambda _{c}^{\ast }(2940)$ is small and it decays almost at rest with {the}
momentum of {the} proton and {the} $D^{0}$-meson in the centre-of-mass
system of 0.42 GeV/c. Such low-momenta particles are almost invisible for
the COMPASS tracking system while energetic $D^{-}$-meson can be easily
detected. So in spite of impossibility to observe the $\Lambda _{c}^{\ast
}(2940)$ decay directly, its production should manifest itself in the
missing mass spectrum.

\begin{figure}[h]
\begin{center}
\includegraphics[scale=0.4]{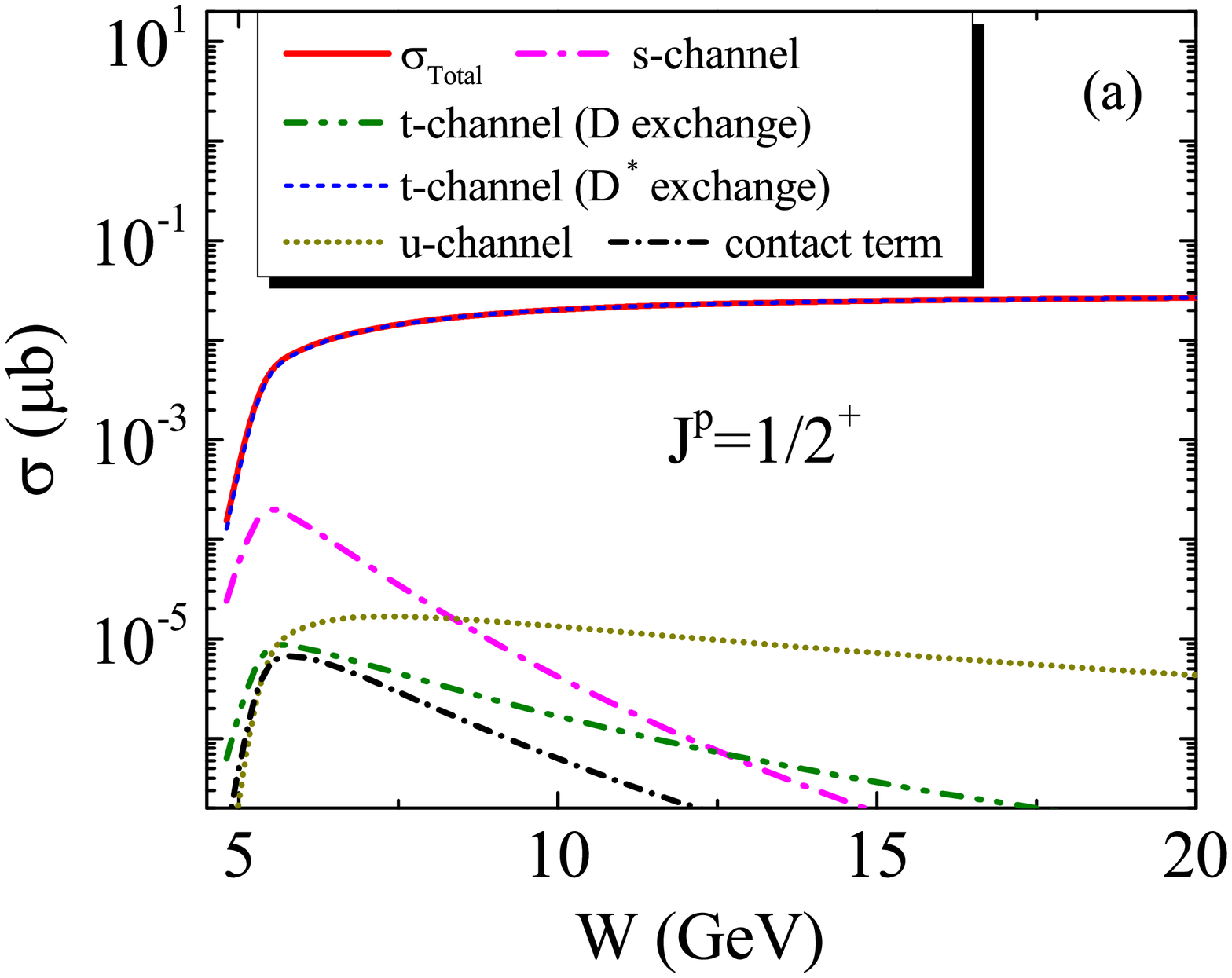} %
\includegraphics[scale=0.4]{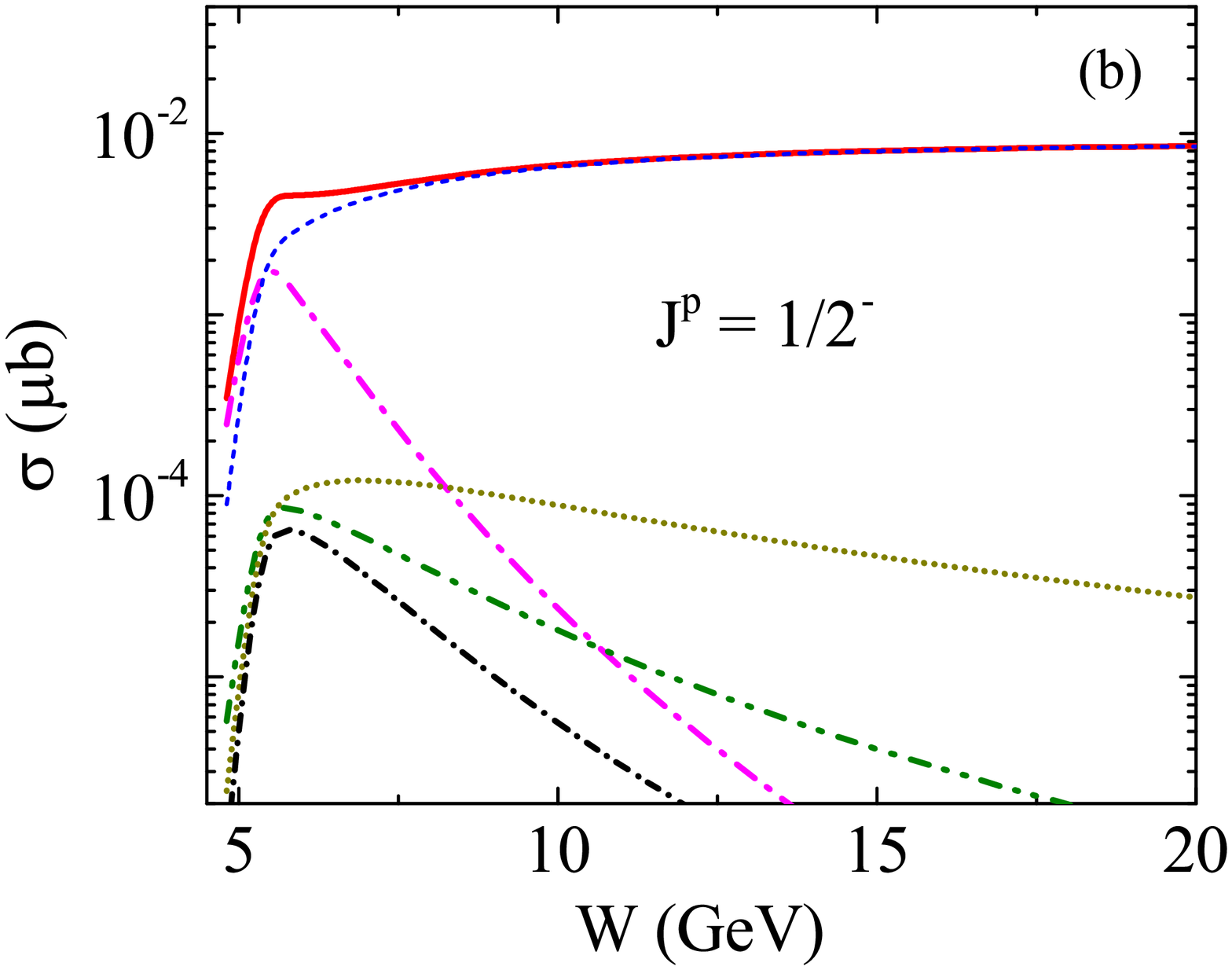}
\end{center}
\caption{(Color online) (a): The total cross section of the $\protect\gamma %
n\rightarrow D^{-}\Lambda _{c}^{\ast +}$ process for the case of $\Lambda
_{c}^{\ast }(2940)$ with $J^{P}=\frac{1}{2}^{+}$. (b) is same as the (a),
but for the case of $\Lambda _{c}^{\ast }(2940)$ with $J^{P}=\frac{1}{2}^{-}$%
. Here, the cross section are calculated with $\Lambda =3.0$ GeV.}
\end{figure}

\section{Summary}

We investigated the photoproduction of {the} charmonium-like $Z_{c}(4200)$ {%
state and the} charmed $\Lambda _{c}^{\ast }(2940)$ baryon. The upper limit
for the partial decay width of $Z_{c}({4200)}\rightarrow J/\psi \pi $ is {%
established using the COMPASS results}, which may provide valuable
information for better understanding of the nature of this particle. The
possibility to observe the production of the charmed $\Lambda _{c}^{\ast
}(2940)$ baryon at the COMPASS conditions is also studied.

The authors acknowledge the COMPASS Collaboration for allowing us to use the
data of the $J/\psi \pi ^{\pm }$ mass spectrum.


\begin{thebibliography}{99}
\bibitem{lx14} X. Liu, Chin. Sci. Bull. 59, 3815 (2014).

\bibitem{sl15} S. L. Olsen, Front. Phys. 10, 101401 (2015).

\bibitem{compass} P.~Abbon \textit{et al}. (COMPASS Collaboration),
Nucl.~Instrum.~Meth. A577, 455 (2007), [arXiv:0703049~[hep-ex]].

\bibitem{a2xy} X. Y. Wang and A. Guskov, arXiv:1510.00898[hep-ph].

\bibitem{belle14} K. Chilikin \textit{et al}. (Belle Collaboration), Phys.
Rev. D 90, 112009 (2014).

\bibitem{zsl14} L. Zhao, W. Z. Deng and S. L. Zhu, Phys. Rev. D 90, 094031
(2014).

\bibitem{wei11} W. Chen and S. L. Zhu, Phys. Rev. D 83, 034010 (2011).

\bibitem{wei15} W. Chen \textit{et al., }Eur. Phys. J. C 75, 358 (2015).

\bibitem{wzg15} Z. G. Wang, Int. J. Mod. Phys. A 30, 15501 68 (2015).

\bibitem{babar07} B. Aubert \textit{et al}. (\textit{BABAR} Collaboration),
Phys. Rev. Lett. 98, 012001 (2007).

\bibitem{belle07} K. Abe \textit{et al}. (Belle Collaboration), Phys. Rev.
Lett. 98, 082001 (2007).

\bibitem{z4200} X. Y. Wang, X. R. Chen and Alexey Guskov, Phys. Rev. D 92,
094017 (2015).

\bibitem{lam2940} X. Y. Wang, Alexey Guskov and X. R. Chen, Phys. Rev. D 92,
094032 (2015).

\bibitem{compassoch} C. Adolph \textit{et al.} (COMPASS Collaboration),
Phys. Rev. D87 (2013) 5, 052018.
\end{thebibliography}
\end{document}